\documentclass[
reprint,
superscriptaddress,
amsmath,amssymb,
aps,
pra,
]{revtex4-1}

\usepackage{graphicx}% Include figure files
\usepackage{dcolumn}% Align table columns on decimal point
\usepackage{bm}% bold math
\usepackage[dvipsnames]{xcolor}
\usepackage[colorlinks=true, citecolor={blue!80!black}, urlcolor={blue!50!black}, linkcolor = {blue!80!black}]{hyperref}
\usepackage{xcolor}
\usepackage{siunitx}
\usepackage{mhchem}
\usepackage{amsmath}
\usepackage{hyperref}

\graphicspath{{figures/}}

\begin{document}
%\preprint{AIP/123-QED}
\title{Destructive Little-Parks Effect in a Full-Shell Nanowire-based Transmon}% 

\author{Deividas~Sabonis}
\thanks{These authors contributed equally to this work}
\affiliation{Center for Quantum Devices, Niels Bohr Institute, University of Copenhagen, 2100 Copenhagen, Denmark}
\affiliation{Microsoft Quantum Lab--Copenhagen, Niels Bohr Institute, University of Copenhagen, 2100 Copenhagen, Denmark}

\author{Oscar~Erlandsson}
\thanks{These authors contributed equally to this work}
\affiliation{Center for Quantum Devices, Niels Bohr Institute, University of Copenhagen, 2100 Copenhagen, Denmark}
\affiliation{Microsoft Quantum Lab--Copenhagen, Niels Bohr Institute, University of Copenhagen, 2100 Copenhagen, Denmark}

\author{Anders~Kringh{\o}j}
\thanks{These authors contributed equally to this work}
\affiliation{Center for Quantum Devices, Niels Bohr Institute, University of Copenhagen, 2100 Copenhagen, Denmark}
\affiliation{Microsoft Quantum Lab--Copenhagen, Niels Bohr Institute, University of Copenhagen, 2100 Copenhagen, Denmark}
 
\author{Bernard~van~Heck}
\affiliation{Microsoft Quantum Lab Delft, Delft University of Technology, 2600 GA Delft, The Netherlands}

\author{Thorvald~W.~Larsen}
\affiliation{Center for Quantum Devices, Niels Bohr Institute, University of Copenhagen, 2100 Copenhagen, Denmark}
\affiliation{Microsoft Quantum Lab--Copenhagen, Niels Bohr Institute, University of Copenhagen, 2100 Copenhagen, Denmark}
 
\author{Ivana Petkovic}
\affiliation{Center for Quantum Devices, Niels Bohr Institute, University of Copenhagen, 2100 Copenhagen, Denmark}
\affiliation{Microsoft Quantum Lab--Copenhagen, Niels Bohr Institute, University of Copenhagen, 2100 Copenhagen, Denmark}

\author{Peter~Krogstrup}
\affiliation{Center for Quantum Devices, Niels Bohr Institute, University of Copenhagen, 2100 Copenhagen, Denmark}
\affiliation{Microsoft Quantum Materials Lab--Copenhagen, 2800 Lyngby, Denmark}

\author{Karl~D.~Petersson}
\affiliation{Center for Quantum Devices, Niels Bohr Institute, University of Copenhagen, 2100 Copenhagen, Denmark}
\affiliation{Microsoft Quantum Lab--Copenhagen, Niels Bohr Institute, University of Copenhagen, 2100 Copenhagen, Denmark}

\author{Charles~M.~Marcus}
\affiliation{Center for Quantum Devices, Niels Bohr Institute, University of Copenhagen, 2100 Copenhagen, Denmark}
\affiliation{Microsoft Quantum Lab--Copenhagen, Niels Bohr Institute, University of Copenhagen, 2100 Copenhagen, Denmark}


\begin{abstract}

A semiconductor transmon with an epitaxial Al shell fully surrounding an InAs nanowire core is investigated in the low $E_J$/$E_C$ regime. Little-Parks oscillations as a function of flux along the hybrid wire axis are destructive, creating lobes of reentrant superconductivity separated by a metallic state at a half-quantum of applied flux. In the first lobe, phase winding around the shell can induce topological superconductivity in the core. Coherent qubit operation is observed in both the zeroth and first lobes. Splitting of parity bands by coherent single-electron coupling across the junction is not resolved beyond line broadening, placing a bound on Majorana coupling, $E_{M}/h\,<\,$10~MHz, much smaller than the Josephson coupling $E_{J}/h\sim4.7$~GHz.
\end{abstract}

\maketitle

Variants of the conventional metallic transmon qubit \cite{koch2007charge,devoret2013superconducting} based on a semiconductor nanowires (NWs) with an epitaxial superconducting shell have appeared recently and shown great promise for qubit applications, offering atomically precise interfaces and electrostatic control of junction properties \cite{de2015realization, larsen2015semiconductor, Casparis2016, hays2018direct, Luthi2018}.  By now, qubit spectroscopy \cite{de2015realization}, coherence \cite{larsen2015semiconductor},  two-qubit operation \cite{Casparis2016}, gate-controlled qubit coupling \cite{Casparis2019} and dc monitoring \cite{Kringhoj2020} have been demonstrated, along with investigations of applied magnetic field \cite{Luthi2018},  junction Andreev bound states \cite{hays2018direct}, anharmonicity \cite{Kringhoj2018}, charge dispersion \cite{bargerbos2019controlling, Kringhoj2019}, spin  \cite{hays2019continuous}, and parity protection \cite{Larsen2020}. Not considered, however, are the consequences of a fully surrounding Al shell used to proximitize the NW core. This is more than a detail. In these small-diameter systems, it is known that an axial magnetic field applied to a cylindrical superconducting shell can give rise to an extreme form of the Little-Parks effect \cite{Little1962} characterized by flux-driven reentrant superconductivity \cite{Liu2001,Sternfeld2011,SolePRB2020}, and, in semiconductor-superconductor hybrids to a possibility to realize topological superconductivity \cite{SoleScience2020}. 

Proposals to realize transmons based on topological superconductors typically require sizeable applied magnetic fields to drive the system into the topological phase \cite{Hassler2011, Hyart2013, Ginossar2014, Ohm2015}. In these schemes, a properly oriented magnetic field splits the spin-orbit band, typically requiring $\sim 1$~T to reach the topological regime \cite{Lutchyn2018}, a challenge to most superconducting technologies \cite{Luthi2018} with some exceptions \cite{kroll}. Recent experiments and theory indicate that topological superconductivity can be induced in a NW with a fully surrounding superconducting shell by winding the superconducting phase around the wire using magnetic flux rather than Zeeman coupling, requiring lower fields ($\sim 0.1$~T) to reach the first reentrant superconducting lobe in the destructive Little-Parks effect regime \cite{SoleScience2020}.

In an ideal topological transmon, a pair of Majorana zero modes (MZMs) straddle the Josephson junction, and are coupled via coherent single-electron tunneling with a characteristic energy scale $E_M$, which accompanies the usual Josephson coupling, $E_{J}$, across the junction \cite{Ginossar2014, Vanzanten2020}. From the perspective of circuit quantum electrodynamics (cQED) spectroscopy, $E_M$ splits the parity-preserving level crossings thereby doubling the number of spectral lines \cite{Ginossar2014}.

In this Letter, we investigate InAs/Al hybrid semiconductor-based NWs with a fully surrounding superconducting shell, comparing qubit spectra and coherence in the zeroth (trivial) lobe to the first reentrant lobe, corresponding to roughly one flux quantum ($\Phi_0 = h/2e$) of applied magnetic field through the Al shell, where topological superconductivity is anticipated.  We focus on the offset-charge-sensitive (OCS) transmon regime \cite{Schreier2008} to facilitate a direct examination of the parity of the superconducting island. Coherence was observed via spectral and time-domain measurements in both the zeroth and first superconducting lobes. The destructive Little-Parks effect causes a field-dependent superconducting gap, and the destruction of superconductivity altogether between lobes, around an applied flux of $\Phi_0/2$. We model the effects of flux on qubit frequency, finding good agreement with experiment. We do not observe clear splittings of gate-modulated charge dispersion bands. Such splitting would reflect coherent Majorana coupling anticrossing the parity bands. This allows us to set a bound on Majorana coupling across the semiconductor junction, $E_{M}/h< 10$~MHz.

\begin{figure}[b]
    \includegraphics[width=0.45\textwidth]{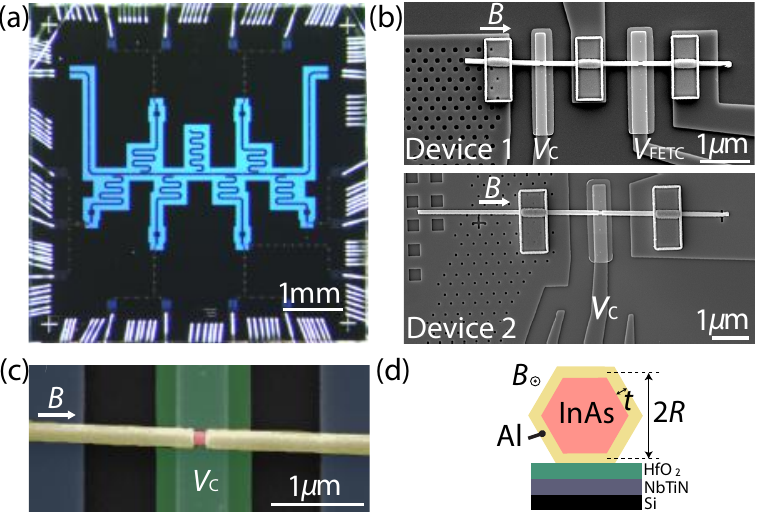}
    \centering
    \caption{(a) Optical micrograph of the sample chip showing multiple separate qubit readout resonators connected to a single readout line. (b) Scanning electron micrographs of two investigated devices: top - device 1, bottom - device 2. Gate voltage $V_C$ in both devices controlled both qubit frequency (large range) and offset charge (small range). Device 1 was decoupled from dc lead (right side) with negative $V_{\mathrm{FETC}}$. The capacitor island is connected in the middle of device 1 and on right side in device 2. (c) Semiconductor junction (red) where Al shell (yellow) is removed, controlled by gate below the dielectric (green). (d) Schematic cross section of the hybrid nanowire and substrate stack, showing mean diameter, $2R$, shell thickness, $t$ and direction of applied field $B$.}
    \label{fig1} 
\end{figure}

A chip containing six NW hybrid transmons and individual readout resonators is shown in Fig.~\ref{fig1}(a). The circuit ground plane, qubit island (with charging energy $E_C/h$~=~512~MHz estimated from electrostatic simulation), and all control and readout elements (electrostatic gates, on-chip gate-filters, readout resonator, transmission line and \ce{NbTiN} crossovers on all control lines) were fabricated from sputtered NbTiN, patterned using e-beam lithography and reactive ion etching on a high-resistivity silicon wafer. The NW is approximately 10~$\mu$m long and resides on top of a \ce{HfO2} dielectric that separates it from the bottom gates. Eight working devices were measured and showed similar results. We present data from two representative devices (device 1 and device 2) shown in Fig.~\ref{fig1}(b). For device 1, one side of the NW is connected to ground plane and the other makes an ohmic contact with NbTiN, allowing dc monitoring that can be connected and disconnected {\it in situ} \cite{Kringhoj2020}. The contact in the middle of the NW connects the junction to the qubit island. Device 2 is a simpler design, with one side of the NW connected to the qubit island and the other to the ground plane. All measurements are performed in a BlueFors XLD dilution refrigerator with a base temperature of 20~mK and 1-1-6 vector magnet.

Figure \ref{fig1}(c) shows the gate-controlled Josephson junction in the full-shell NW. The junction was formed by wet-etching the Al shell in a $100-150$~nm region defined by electron-beam lithography. The voltage $V_C$ on the bottom gate was used to tune the junction coupling. The same gate was used to tune both the qubit frequency, on a large voltage range, and the offset charge, on a small voltage range. The schematics of the cross section of the full-shell InAs/Al NW together with a substrate material stack is shown in Figure \ref{fig1}(d).  

\begin{figure}[b]
    \includegraphics[width=0.5\textwidth]{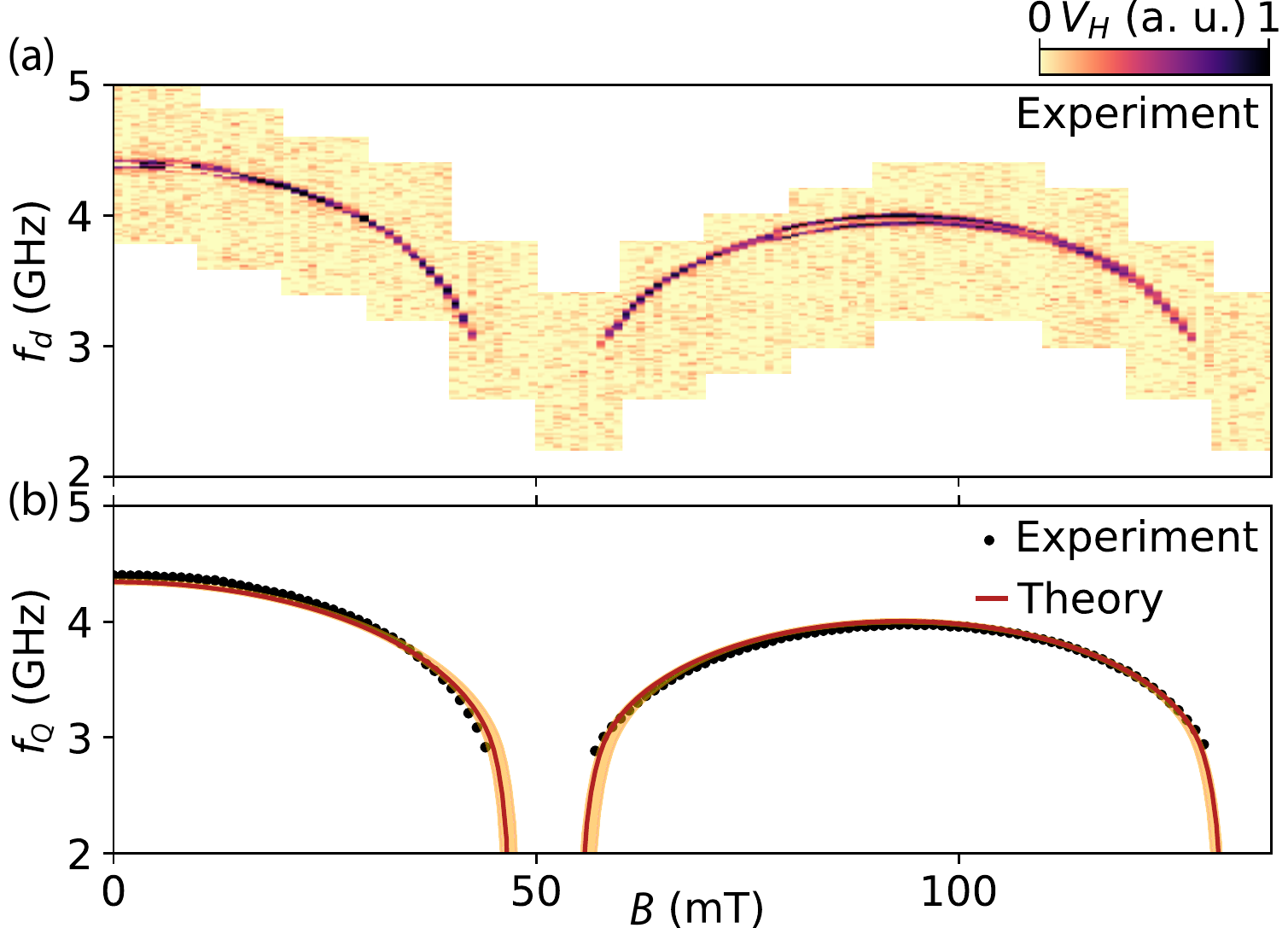}
    \centering
    \caption{(a) Two-dimensional map of two-tone spectroscopy as a function of drive frequency ($f_d$) and parallel magnetic field $B$ for device 1. Four regimes are seen: (i) Low fields ($0-40$~mT), zeroth superconducting lobe, qubit frequency $f_{Q}$ decreases with increasing $B$; (ii) Destructive regime, around half flux quantum $\Phi_{0}/2\sim 50$~mT, qubit coherence is lost; (iii) Higher fields ($60-120$~mT), first lobe, coherence is recovered with slightly reduced $f_{Q}$; (iv) Second destructive regime at $\sim 135$~mT where coherence is again lost. (b) Experiment (black points) and theory (orange curve) for qubit frequency, $f_{Q}$. Width of theory curve marks 3rd and 97th percentiles of the fit.}
        \label{fig2}
\end{figure}

As a first signature of the Little-Parks lobe structure, we examine qubit frequency as a function of axial magnetic field, $B$. The zero-field qubit frequency was set to $\sim 4.3$~GHz ($V_C\sim -2.5$~V) slightly below the readout resonator frequency, $f_r\sim 5.29$~GHz for device 1. A map of the amplitude of the heterodyne demodulated transmission $V_H$ (line average subtracted) measured in two-tone spectroscopy as a function of qubit drive frequency $f_d$ (2~$\mu$s long drive pulse followed by a readout tone) and magnetic field $B$ is shown in Fig. \ref{fig2}(a). At each field step the readout frequency was adjusted by tracking the frequency of the resonator. At low fields, $B\lesssim 40$~mT, the NW is in the zeroth lobe, with a trivial proximity effect from the shell, and a field dependent gap, $\Delta(B)$ with its maximum value, $\Delta_{0}$ at $B = 0$. At intermediate fields, around $\sim 50$~mT, the qubit frequency falls rapidly, the resonance broadens and disappears. This behavior is associated with the destructive regime around applied flux $\Phi_0/2$, where $\Delta(B)$ goes to zero and superconductivity is destroyed \cite{SolePRB2020, SoleScience2020}. At higher fields, $B\sim 60-120$~mT the qubit is recovered and shows similar behavior to the low-field regime with a maximum qubit frequency around 90~mT, slightly reduced from its zero-field value. This recovery of qubit frequency reflects the underlying lobe structure of superconductivity in the NW  \cite{SolePRB2020}. In the first lobe, one vortex has entered the shell, the superconducting phase twists once around the wire circumference, and topological superconductivity can be realized in the core \cite{SoleScience2020}.

We model the field dependence of the qubit frequency by considering a cylindrical shell of radius $R$ and thickness $t$ in an axial magnetic field $B$. Following Refs.~\cite{Sternfeld2011,dao2009destruction,SolePRB2020}, the pair-breaking term $\alpha (B)$ is calculated as 

\begin{equation}
\alpha(B) = \frac{4\xi^2 k_B T_{C0}}{\pi R^2} \bigg[ \bigg(n-\frac{\Phi}{\Phi_0}\bigg)^{2} +\frac{t^2}{4R^2} \bigg( \frac{\Phi^2}{\Phi_0^{2}} +\frac{n^2}{3} \bigg) \bigg],
\end{equation}
where $\xi$ is the zero-field coherence length, $T_{C0}$ is the zero field critical temperature and $\Phi$ = $\pi R^2 B$ is the applied magnetic flux. The winding number $n$ which is energetically favourable is determined by minimising $\alpha$ versus $n$ at a given magnetic field $B$. Once $\alpha (B)$ is known, the field dependence of the pairing energy $\Delta (B)$ can be determined numerically \cite{Larkineq26} based on Larkin's theory for cylindrical superconductors in a magnetic field \cite{larkin1965superconductor}. The expected scaling of the qubit frequency, $f_{Q}=f_{0}\,d(B)^{1/2}$ used (assuming $f_{Q}\sim \sqrt{E_J/h}$ \cite{koch2007charge} and $E_J \sim \Delta$ \cite{tinkham2004introduction}), where $d(B)=\Delta(B)/\Delta_0$ was then used to determine the four fit parameters, $R=81$~nm, $t=36$~nm, $\xi=139$~nm, and $f_{0}=4.34$~GHz, yielding the theory curve in Fig.~\ref{fig2}(b), which is in good agreement with experiment. The extracted geometrical fit parameters $t$ and $R$ are comparable to the values estimated from the electron micrograph in Fig.~\ref{fig1}(b). Details of the fitting procedure are given in the in an accompanying analysis notebook \cite{zenodo}.

\begin{figure}[t]
    \includegraphics[width=0.48\textwidth]{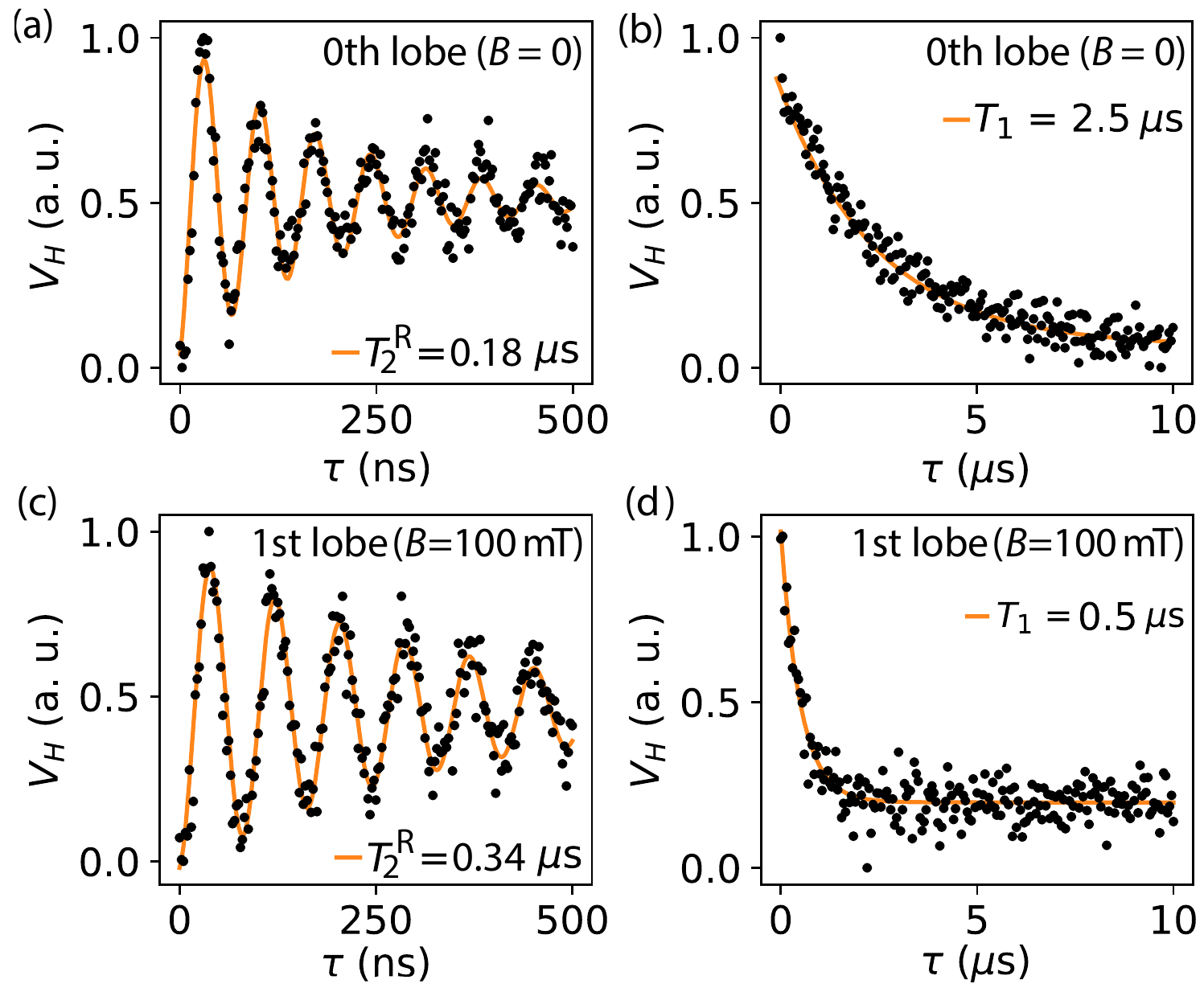}
    \centering
    \caption{(a) Rabi oscillations (black points) in the zeroth lobe ($B=0$) in device 1 with best fit exponentially decaying sinusoidal oscillation, $A\exp(-\tau/T_{2}^{R}) \sin(\omega\tau+\phi)+C$ (orange) as a function of pulse duration $\tau$, yields Rabi time $T_{2}^R=0.34\,\mu$s at qubit frequency $f_{Q}=4.53$~GHz. (b) Qubit relaxation (black points) as a function of delay $\tau$, with best fit to exponential decay, $A \exp(-\tau/T_1)+C$ (orange) yields $T_1=2.5\,\mu$s. (c, d) Same as (a, b) except in the first lobe ($B=100$~mT) yield Rabi and relaxation times, $T_{2}^R=0.18\,\mu$s and $T_1=0.5\,\mu$s at qubit frequency $f_{Q}=3.75$~GHz.}
        \label{fig3}
\end{figure}

Time-domain measurements in the zeroth and first lobes are compared in Fig.~\ref{fig3}. In the zeroth lobe ($B =0$), Rabi oscillations with qubit drive frequency $f_d$ set to match  $f_Q$~=~4.53~GHz yield a Rabi decay time of $T_{2}^R$~=~0.18~$\mu$s. Qubit lifetime in the zeroth lobe was measured by exciting the qubit using the $\pi$ pulse, the length of which is found from Fig.~\ref{fig3}(a), and measuring the qubit response after a time $\tau$, yielding $T_1$~=~2.5~$\mu$s.  In the first lobe ($B=100$~mT), the Rabi decay time $T_{2}^R$~=~0.34 $\mu$s was measured with relaxation time $T_1$~=~0.5 $\mu$s at $f_Q$~=~3.75~GHz somewhat shorter than in the zeroth lobe, which we speculate could be attributed to dissipation from subgap states present in the first lobe \cite{SoleScience2020}. No obvious features indicating discrete subgap or zero-energy states in the first lobe were seen in time-domain measurements.

\begin{figure}[b]
    \includegraphics[width=0.45\textwidth]{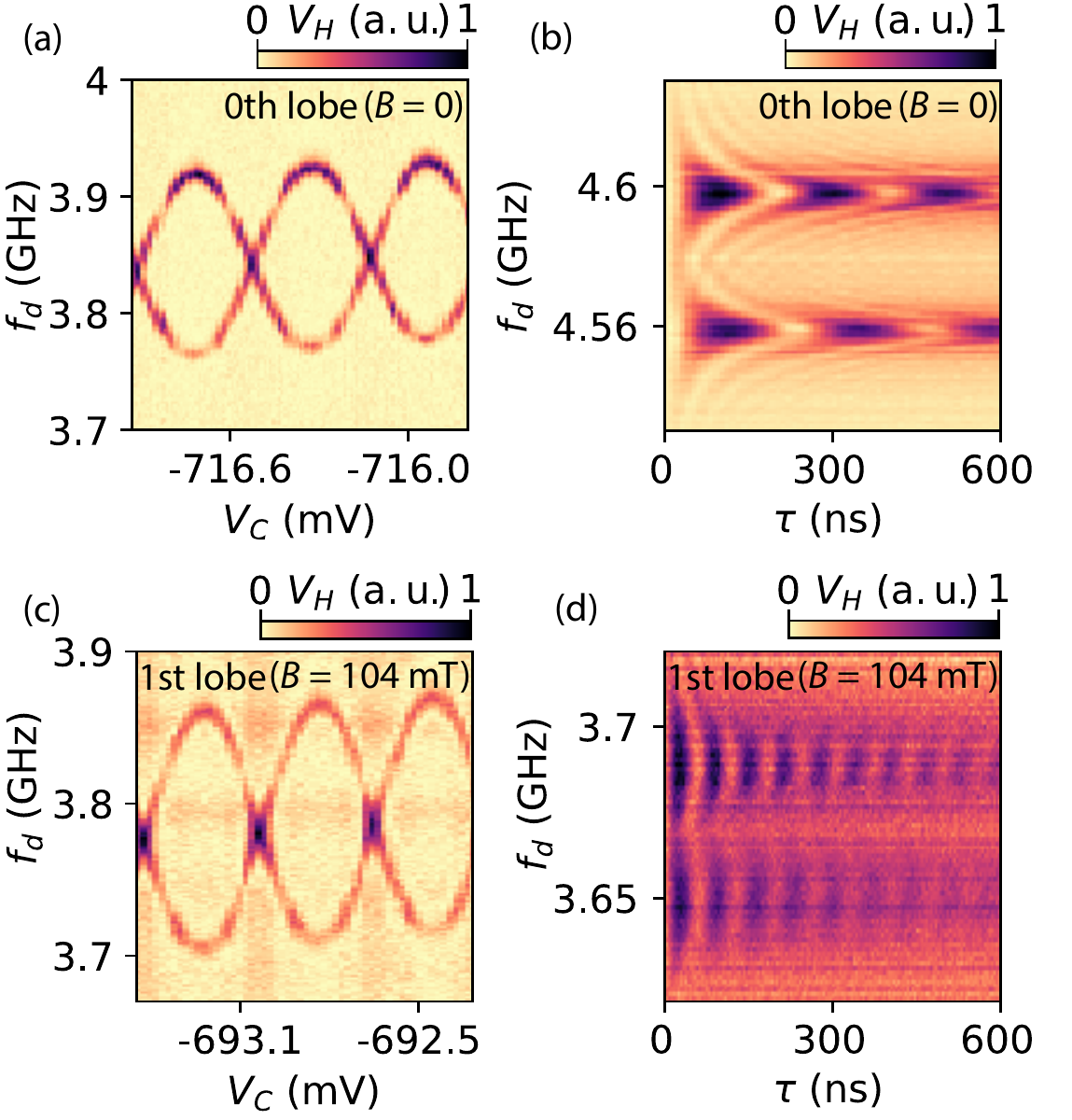}
    \centering
    \caption{(a) Qubit spectrum as a function of the gate voltage $V_C$ in the zeroth lobe ($B=0$) for device 2 showing two sinusoidal parity branches. The simultaneous appearance of both branches indicates that parity switching (poisoning) is faster than data acquisition time. (b) Coherent oscillations of both parity branches at slightly different $V_C$ compared to (a) as a function of the drive time $\tau$ in the zeroth lobe ($B=0$). (c, d) Same as (a, b) except in the first lobe ($B$~=~104~mT). Higher frequency oscillations in first lobe (d) reflects higher effective drive power than in (b).}
    \label{fig4}
\end{figure}

We next examine charge dispersion in the two lobes in device 2. The fine-scale dependence of the qubit transition frequency on gate voltage $V_C$, swept over a small range, is shown in Fig.~\ref{fig4}(a). The interlaced sinusoidal qubit response as a function of $V_C$ (column average subtracted) represents the two charge parity branches, visible in the OCS regime, as investigated previously in conventional~\cite{Schreier2008} and semiconducting junctions \cite{bargerbos2019controlling,Kringhoj2019}. The simultaneous visibility of both parity branches indicates parity switching (poisoning) that is fast compared to the measurement time \cite{Schreier2008}. 

Coherent qubit oscillations in both parity branches in the zeroth lobe ($B=0$) are shown in Fig.~\ref{fig4}(b), (qubit response $V_H (\tau=0)$ subtracted from all columns for better visibility) measured at more positive $V_C$, which increased the qubit frequency and slightly reduced dispersion. Rabi oscillations as a function of qubit drive time $\tau$ and drive frequency, $f_d$, are similar in the two parity branches. Comparable measurements in the first lobe ($B =104$~mT), are shown in Fig.~\ref{fig4}(c, d). Overall, no clear splitting at the parity crossing can be resolved here or in a broader range of gate voltages. We are able to place an upper bound, $E_{M}/h<$ 10~MHz \cite{zenodo} based on the absence of a well-resolved splitting. We do not know if the absence of resolvable splitting of the parity branches reflects an absence of zero-energy states, a lack of coupling across the junction, or a short coherence due to rapid poisoning \cite{Luthi2018,Serniak2018} in the first lobe. 
\section*{Acknowledgments}

We thank Marina Hesselberg, Karthik Jambunathan, Robert McNeil, Karolis Parfeniukas, Agnieszka Telecka, Shivendra Upadhyay, and Sachin Yadav, for device fabrication. We also thank Lucas Casparis, Ruben Grigoryan, Eoin O'Farrell, Saulius Vaitiek\.enas, Judith Suter, and David van Zanten for valuable discussions. Research was supported by Microsoft, the Danish National Research Foundation, and the European Research Council under grant HEMs-DAM No.716655.

\bibliography{LittleParkscQED3}% Produces the bibliography via BibTeX.

\begin{thebibliography}{33}
\bibitem [{Lar()}]{Larkineq26}%
  \BibitemOpen
  \href@noop {} {\bibinfo  {journal} {In detail, the following implicit
  equation need to be solved for $\Delta$ given $\alpha$:
  $\ln{\frac{\Delta_0}{\Delta}}$~=~$\frac{\pi\alpha}{4\Delta}$ for
  $\alpha>\Delta$ and
  $\ln{\frac{\Delta_0}{\Delta}}$~=~$\ln{\frac{\alpha+(\alpha^2-\Delta^2)^{1/2}}{\Delta}-\frac{(\alpha^2-\Delta^2)^{1/2}}{2\alpha}}$+$\frac{\alpha}{2\Delta}\arcsin{\frac{\Delta}{\alpha}}$
  for $\alpha<\Delta$. This expression appears as Eq.~26 in Ref.~[30]}\
  }\BibitemShut {NoStop}%
\end{thebibliography}

\end{document}